\documentclass[pdflatex,sn-mathphys-num,iicol]{sn-jnl}

\usepackage{graphicx}%
\usepackage{multirow}%
\usepackage{tabularx}%
\usepackage{amsmath,amssymb,amsfonts}%
\usepackage{amsthm}%
\usepackage{mathrsfs}%
\usepackage[title]{appendix}%
\usepackage{xcolor}%
\usepackage{textcomp}%
\usepackage{manyfoot}%
\usepackage{booktabs}%
\usepackage{algorithm}%
\usepackage{algorithmicx}%
\usepackage{algpseudocode}%
\usepackage{listings}%

\theoremstyle{thmstyleone}%
\theoremstyle{thmstyletwo}%

\theoremstyle{thmstylethree}%

\begin{document}

\title[Universal scaling and relaxation in decaying turbulence of Bose gases]{Universal scaling and relaxation in decaying turbulence of Bose gases}


\author*[1]{\fnm{Arnol D.} \sur{Garc\'ia-Orozco}}\email{arnolgarcia@ifsc.usp.br}

\author[1]{\fnm{Michelle A.} \sur{Moreno-Armijos}}\email{michelle.moreno@ifsc.usp.br}
\equalcont{These authors contributed equally to this work.}

\author[1]{\fnm{Sarah} \sur{Sab}}\email{sarahsab@usp.br}
\equalcont{These authors contributed equally to this work.}

\author[1]{\fnm{Amilson} \sur{Fritsch}}\email{amilson@usp.br}
\equalcont{These authors contributed equally to this work.}

\author[1,2,3]{\fnm{Vanderlei} \sur{Salvador Bagnato}}\email{vander@ifsc.usp.br}
\equalcont{These authors contributed equally to this work.}

\affil*[1]{\orgdiv{Instituto de F\'isica de S\~ao Carlos}, \orgname{Universidade de S\~ao Paulo}, \orgaddress{\city{S\~ao Carlos}, \postcode{13566-590}, \state{S\~ao Paulo}, \country{Brazil}}}

\affil[2]{\orgdiv{Department of Biomedical Engineering}, \orgname{Texas A\&M University}, \orgaddress{\city{College Station}, \postcode{77843}, \state{Texas}, \country{USA}}}

\affil[3]{\orgdiv{Hagler Institute for Advanced Study}, \orgname{Texas A\&M University}, \orgaddress{\city{College Station}, \postcode{77843}, \state{Texas}, \country{USA}}}

\abstract{
Understanding the relaxation dynamics of isolated quantum systems remains a central challenge in nonequilibrium physics. In this short review, we discuss recent experimental and theoretical advances in universal dynamics of turbulent Bose gases, with particular emphasis on nonthermal fixed points and wave turbulence in our experimental system at the São Carlos Institute of Physics of the University of São Paulo. We highlight the emergence of self-similar scaling behavior and its connection to particle and energy transport across momentum scales. In addition, we discuss an approach in which a differential equation has a universal scaling solution, providing a practical framework for extracting universal exponents from limited regions of the momentum distribution. Furthermore, we recap experimental observations of direct and inverse cascades in a three--dimensional trapped Bose--Einstein condensate, revealing distinct relaxation stages governed by universal scaling laws. These results demonstrate that turbulence plays a key role in far--from--equilibrium dynamics, offering a unified perspective on transport processes and thermalization in quantum many--body systems.
}


\maketitle

\section{Introduction}\label{sec:introduction}
The study of nonequilibrium dynamics in closed and open systems, and the routes to thermalization, is of fundamental importance across several areas of physics, ranging from cosmology \cite{Kofman1994,Micha2004,Berges2008,Allahverdi2010} to condensed matter \cite{Polkovnikov2011,Calabrese2006,Eisert2015,Mitra2018} and quantum many-body systems \cite{Deutsch1991,Srednicki1994,Rigol2008,Dalessio2016}. In this context, a central question is how an isolated system evolves after being driven out of equilibrium and whether its dynamics can be described by universal behavior. Similar to equilibrium critical phenomena, far-from-equilibrium systems can be classified into universality classes based on their dynamical properties \cite{Widom1965,Kadanoff1966,Wilson1971a,Wilson1971b,Hohenberg1977,Janssen1979}. However, in this situation, time serves as a relevant scaling parameter, particularly when the system is driven far from equilibrium \cite{Hohenberg1977,Janssen1979}.

Quantum turbulence provides a natural framework to investigate these questions in quantum fluids. In systems such as Bose--Einstein condensates (BECs), turbulence can be generated through different mechanisms, including the decay of multiple quantum vortices and the nonlinear evolution of density fluctuations \cite{Henn2009TurbulencePRL,Henn2009TurbulencePRA,Madison2001Vortices}. In both situations, the system develops a turbulent state characterized by a power-law behavior in the momentum distribution \cite{Thompson2014TurbulenceBEC,Navon2016TurbulentCascade}. This behavior indicates the presence of transport processes across momentum scales and suggests the emergence of universal scaling laws.

Most recently, the concept of nonthermal fixed points (NTFPs) has been introduced as a framework to describe universal dynamics in far-from-equilibrium systems \cite{BergesNuclearPhysicsB2009,Mikheev2023,NowakPRA2012}. Within this approach, the evolution of relevant observables, such as the momentum distribution, exhibits self-similar scaling in time and momentum \cite{Mikheev2023,Madeira2022Review,Berges2015,Chantesana2019}. This scaling behavior is associated with transport processes, such as direct and inverse cascades, which redistribute particles and energy across different momentum scales. Therefore, ultracold atomic gases provide an ideal experimental platform due to their high level of control and the possibility to probe their dynamics with high precision \cite{Erne2018,Prufer2018,Glidden2021,GarciaOrozcoPRA2022}.

In this review, we discuss recent experimental and theoretical advances in the study of universal dynamics in turbulent Bose gases, with particular emphasis on our experimental system at the São Carlos Institute of Physics (IFSC--USP). We focus on the emergence of scaling behavior, the role of nonthermal fixed points, and the connection between turbulence and transport processes. The paper is organized as follows. In Sec. \ref{sec:ntfp_wwt}, we introduce the concept of nonthermal fixed points and weak wave turbulence. In Sec. \ref{sec:universal_ntfp}, we discuss universal scaling and its connection to conservation laws. In Sec. \ref{sec:experimental}, we present the experimental system used in our study. In Sec. \ref{sec:universal_tbg}, we describe the observation of universal dynamics in turbulent Bose gases. In Sec. \ref{sec:diferencial_equation}, we introduce a differential-equation approach for extracting universal exponents. In Sec. \ref{sec:transport_turbulence}, we discuss the implications for turbulence and transport. Finally, in Sec. \ref{sec:outlook}, we present an outlook. 

\section{Nonthermal fixed points and weak wave turbulence}\label{sec:ntfp_wwt}
The concept of NTFPs in non-equilibrium systems was introduced by Jürgen Berges and Gabriele Hoffmeister in 2008 \cite{BergesNuclearPhysicsB2009}. They present this as a basic property of quantum field theories. The NTFP theory is based on renormalization-group concepts and can be applied to various systems in physics \cite{BergesPRL2008,Micha2004,Zakharov1992,Schmied2019}. In the specific case of ultracold atoms and many-body dynamics, Thomas Gasenzer's group \cite{Mikheev2023EPJSP,NowakPRA2012,Mikheev2023,Schmied2019} has made significant contributions to the development of the theory, and other groups around the world have studied this theory using experimental systems such as ultracold atoms and BEC. 

Let us consider a dilute, degenerate Bose gas that is compressible, meaning that collective sound-wave excitations can occur. In this situation, the evolution of the Bose gas can be described by the weak kinetic equation (WKE), as collisions between different wave modes dominate the system \cite{NowakPRA2012}. This system has non-trivial stationary solutions that are nonthermal and exhibit power-law behavior in the occupation modes in momentum space. This characteristic in the occupation modes of momentum implies an energy flux from large to small scales; during this flux, the energy passes through the inertial region. In the inertial region, momentum distribution is stationary, and a power law describes its behaviour with a universal exponent predicted by the weak wave turbulence (WWT) theory \cite{Zakharov1992,Nazarenko2011Wave}.

The nonthermal solution for this situation, in which the occupation number $n(k)$ obeys a scaling law in the form, 

\begin{equation}
    n(k) \sim k^{-\zeta},
    \label{eq:ntfp_solution}
\end{equation} 

\noindent where $k$ is the momentum, $\zeta$ is the exponent that depends on the kind of interaction (three--wave, four--wave, etc.) and the kind of flux (particle or energy). However, this solution is not valid for all values of momentum; the WKE fails for the long-wavelength  (infrared--IR) regime when applied to a degenerate Bose gas. This happens because where single--particle occupancy amplitudes grow large, and the description in terms of elastic collisions, for example, two--to--two, becomes inaccurate. Therefore, strong turbulence is expected in the IR regime.

Several works \cite{ScheppachPRA2010,BergesPRL2008,BergesNuclearPhysicsB2009,BergesPRD2011} proposed a new approach to scalability that involves the short-wavelength (ultraviolet -- UV) regime, where the WKE and the Boltzmann equation are valid, and the IR regime. In previous works, the solution proposed can be applied to Bose gases, where the IR regime dominates the system's behavior; using this approach, the solution is known as NTFPs.

In the UV limit, outside the inertial range, energy is coupled from an external or internal source and cascades to high momenta. In contrast, within the inertial range, the flux is continuous across momenta. This process can be described by a continuity equation in momentum space, with a momentum--independent, radially oriented current vector. The single--particle occupation number spectrum has the exponent associated with the number of spatial dimensions of the system $d$ \cite{NowakOxford2016}. For particle flux,

\begin{equation}
    \zeta_{P}^{UV}=d-2/3,
    \label{eq:zeta_particle_UV}
\end{equation}

\noindent and energy flux,

\begin{equation}
    \zeta_{E}^{UV}=d,
    \label{eq:zeta_energy_UV}
\end{equation}

\noindent where the subscripts $P$ and $E$ denote particle and energy flux, respectively.

On the other hand, in the IR limit of long-wavelength, the incompressible superfluid component of the gas dominates the system. The power law in the IR regime arises because of the radial decay of the flow velocity around the vortex core. The single--particle occupation number spectrum shows the following exponent for the power law \cite{ScheppachPRA2010,DosSantosFilhoPhysicsD2022,NowakPRA2012}, for particle flux,

\begin{equation}
    \zeta_{P}^{IR}=d+2,
    \label{eq:zeta_particle_IR}
\end{equation}

\noindent and energy flux,

\begin{equation}
    \zeta_{E}^{IR}=d+2+z,
    \label{eq:zeta_energy_IR}
\end{equation}

\noindent where $z$ is the dynamical scaling exponent for the dispersion relation $\omega(k) \sim k^{z}$.

\section{Universal scaling}\label{sec:universal_ntfp}
The concepts of universality and scaling were first introduced in 1965 by Widom, Kadanoff, and Wilson \cite{Widom1965,Kadanoff1966,Wilson1971a,Wilson1971b} and were quickly generalised to the dynamic case in 1977 \cite{Hohenberg1977,Janssen1979}. In this context, when a physical system is studied within the framework of renormalization group theory \cite{Mikheev2023EPJSP} it can be examined at different resolutions, in particular near a critical point. Near this point, the physical system exhibits self-similarity, meaning it does not change appearance as resolution varies.

To better understand the concept of universal scaling, Gasenzer \textit{et al.} \cite{Mikheev2023EPJSP} introduced a simple example: consider a two-point correlation function $C(x,s)$ of some locally measurable observable, which depends only on the distance $x = |\mathbf{r}_{1}-\mathbf{r}_{2}|$ between two positions $\mathbf{r}_{1}$ and $\mathbf{r}_{2}$ in space, for example, $n(k,t)$ in an atomic system. If the system is homogeneous and isotropic, the parameter $s$ defines the resolution in units of a fixed length scale and represents the renormalization-group flow parameter. As the value of $s$ changes, the correlation function $C(x,s)$ should change accordingly. Self-similarity means that $C(x, s)$ rescales as,

\begin{equation}
    C(x,s) = s^{\delta}f(x/s).
    \label{eq:correlation_scaling}
\end{equation}

Consequently, the correlation function would be characterized only by a universal exponent $\delta$ and a scale function $f$.

The theory of NTFPs is conceived as an analogy to renormalization-group theory in equilibrium critical phenomena \cite{BergesNuclearPhysicsB2009,Mikheev2023EPJSP} which also predicts dynamical scaling in an out-of-equilibrium system near an NTFP. This theory was developed to apply to different physical systems, each governed by distinct microscopic mechanics that dictate their time evolution. Despite their differences, universal dynamical scaling emerges in the relevant quantity $f(x,t)$ that can be expressed in the form,

\begin{equation}
    f(x,t)=t^{\alpha}F(t^{\beta}x),
    \label{eq:f_scaling}
\end{equation}

\noindent where $F$ represents a universal function and $\alpha$ and $\beta$ are universal exponents characterizing the scaling behavior, the two values are nonzero, and are related to amplitude adjustments and scale renormalizations, respectively. In the context of a time-dependent momentum distribution, $n(k,t)$, Eq. \ref{eq:f_scaling} can be rewritten as,
	
\begin{equation}
    n(k,t) =\left(\frac{t}{t_0}\right)^{\alpha}F\left[\left(\frac{t}{t_0}\right)^{\beta} k\right],
    \label{eq:scaling}
\end{equation}

\noindent where $t_0$ is a reference time. The signs of $\alpha$ and $\beta$ indicated the direction of the transport of particles and energy in the system. It is assumed to depend on the modulus k = $|\mathbf{k}|$ of the momentum.

\subsection{Conservation laws}\label{subsec:conservation_laws}
Another important aspect of the NTFPs theory is the conservation law for the transport of particles and energy, given that the theory was developed for use in different physical systems, implying that the universal scaling depends on the $d$-dimensional.

To characterize this transport/flux of particles and energy, two observables are introduced: the average particle number $\overline{N}$ and the kinetic energy $\overline{M}_{2}$. Using Eq. \ref{eq:scaling}, it is possible to express their time evolution in terms of the spatial dimension $d$ of the system and the scaling exponents:

\begin{equation}
    \overline{N}=\int \text{d}^{d}k \, n(k,t)\propto \left( \frac{t}{t_{0}}\right)^{\alpha-d\beta},
    \label{eq:n}
\end{equation}

\begin{equation}
    \overline{M}_{2}=\int \text{d}^{d}k \, k^{2}\frac{n(k,t)}{\overline{N}}\propto\left(\frac{t}{t_{0}}\right)^{-2\beta},
    \label{eq:m2}
\end{equation}

\noindent where the values of the exponents, $\alpha$ and $\beta$, indicate the flux direction of particles or energy in the system.The low and high momentum ranges give the integral limits $(t/t_{0})^{-\beta}k_{l}\leq|k|\leq (t/t_{0})^{-\beta}k_{h}$. 

\section{Experimental systems}\label{sec:experimental}
Ultracold atom systems are often used to study out-of-equilibrium phenomena due to their versatility, high precision, and ability to probe quantum turbulence \cite{Henn2009TurbulencePRL,Neely2013}, topological defects (vortices, solitons) \cite{Khaykovich2002}, quenches \cite{Hung2011,Makotyn2014}, cosmological analogs\cite{Weiler2008}, and, most recently, NTFPs \cite{NowakPRA2012,Nowak2011}. Notably, several groups have reported evidence of NTFP in ultracold Bose gases using different experimental approaches. These include one-dimensional gases studied \cite{Erne2018}, spinor condensates investigated \cite{Prufer2018}, homogeneous three-dimensional systems explored \cite{Glidden2021}, and turbulent trapped condensates analyzed by our group in São Carlos \cite{GarciaOrozcoPRA2022}. These experiments demonstrate the universality of far--from--equilibrium dynamics across different geometries (potentials) and interaction regimes.

\begin{figure}[h!]
    \centering
    \includegraphics[width=\linewidth]{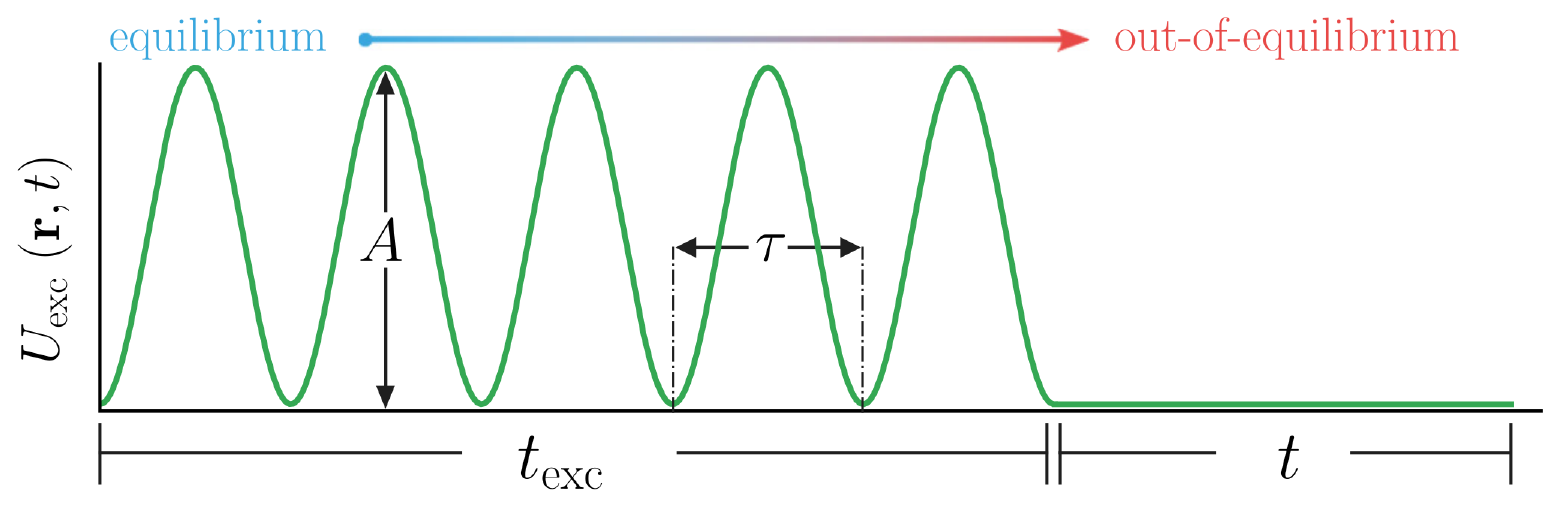}
    \caption{A controllable time-varying magnetic field potential with amplitude A drives the equilibrium BEC to an out-of-equilibrium state. The drive potential is applied for a time $(t_{\mathrm{exc}})$ for around $t_{\mathrm{exc}}=5\tau$ , then turned off. The system evolves in the trap for a time ($t$) without an external potential. After this time, the trap potential is turned off, and after $30\,\mathrm{ms}$ of time of fligth $(t_{\mathrm{TOF}})$, the absorption imaging is taken. Reproduced from Ref.~\cite{GarciaOrozcoPRA2022}.}
    \label{fig:exp_seq}
\end{figure}

In our experimental system, we can reach a BEC (atomic $^{87}$Rb) in two different states, $\vert\mathrm{F=2,m_{F}=2}\rangle$ and $\vert\mathrm{F=1,m_{F}=-1\rangle}$ with number of atoms around $3.5(3)\times10^{5}$ atoms in the ground state $\vert2,2\rangle$ or $2.0(3)\times10^{5}$ atoms $\vert1,-1\rangle$ in a quadrupole-Ioffe configuration (QUIC) magnetic trap \cite{GarciaOrozco2023Thesis,MorenoArmijos2025Thesis}. After achieving BEC equilibrium, we begin the excitation protocol as shown in Fig. \ref{fig:exp_seq}.

The potential to excite the equilibrium BEC is provided by an additional pair of coils in an anti-Helmholtz configuration, rotated by a small angle ($\sim5^{\circ}$) with respect to the principal axis of the QUIC trap \cite{Henn2009TurbulencePRL}. The application of this potential $U_\mathrm{exc}(\mathbf{r},t)$ corresponds to an effective 3D rotation and distortion of the original trap potential shape \cite{Henn2009TurbulencePRA,Madison2001Vortices}. In the experiment, we apply an oscillating current in the coils with frequency close to the radial frequency $\Omega \approx \omega_{r}$ where $\tau=2\pi/\Omega$, and an amplitude $A$ that can be varied from $0$ to $3\mu_{0}$, where $\mu_{0}$ is the chemical potential of the initial equilibrium BEC. For a specific condition, the $f_{\mathrm{exc}}$ and $A$ are fixed, and the hold time $t$ varies to study the dynamics in the trap after excitation.

In our experiment, the atoms are detected using standard absorption images in time of flight (TOF), which give the density distribution of the cloud, $n(r)$. During the expansion, the momentum distribution of each particle is approximately conserved; thus, the density distribution after expansion converges to the in-situ momentum distribution. Therefore, the momentum distribution in the momentum space $n(k)$ is obtained from the momentum distribution in the space $n(r)$ by defining,

\begin{equation}
    k \equiv \frac{m_{\mathrm{Rb}}r}{\hbar t_{\mathrm{TOF}}},
    \label{eq:k_tTOF}
\end{equation}

\noindent where $m_{\mathrm{Rb}}$ is Rubidium mass, $t_{\mathrm{TOF}}$ is the time of flight, that is, the time we wait after turning off the confining potential to take the picture of atoms, and $\hbar$ is reduced Planck's constant. 

This method is well established in the literature and has been successfully used to extract the momentum distribution of turbulent trapped BECs \cite{Thompson2014TurbulenceBEC,Navon2016TurbulentCascade}. In the next sections, we discuss experiments on NTFP in our BEC in a harmonic magnetic trap at IFSC-USP. 

\section{Universal dynamics in turbulent Bose gases}\label{sec:universal_tbg}
The study of universal dynamics in turbulent Bose gases shows a direct connection between quantum turbulence and NTFP. In this situation, demonstrating that a BEC driven out of equilibrium in a harmonic trap exhibits self-similar scaling behavior characteristic of NTFPs \cite{GarciaOrozcoPRA2022}.

Once our system is driven out of equilibrium, we cease the excitation and let the turbulent cloud evolve in the trap. This system dynamics is recorded, and the resulting momentum distribution is analyzed in the context of the universality of dynamics exhibited by far-from-equilibrium quantum systems \cite{Prufer2018,Erne2018,Madeira2022Review,Berges2015,NowakOxford2016,Chantesana2019}. 

\begin{figure}[h!]
    \centering
    \includegraphics[width=0.9\linewidth]{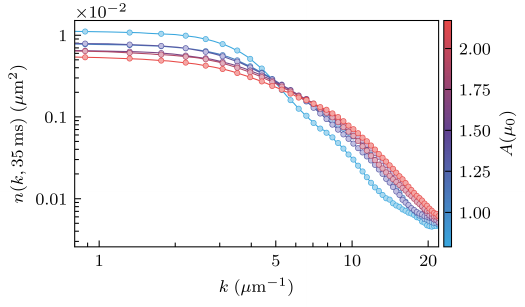}
    \caption{Momentum distribution $n(k,t=35\,\mathrm{ms})$ for different excitation amplitudes. The turbulent state in the cloud is established. The power-law behavior can be observed in the momentum range $10\,\mu\mathrm{m^{-1}} \leq k \leq 17\,\mu\mathrm{m^{-1}}$. Reproduced from Ref.~\cite{GarciaOrozcoPRA2022}.}
    \label{fig:PowerLaw_nkt}
\end{figure}

The first indicator that our system is out of equilibrium is the emergence of power-law behavior in the momentum distribution $n(k,t)$. This helps us identify the type of process the system exhibits. In our experiment for an excitation amplitude around $1.8\mu_{0}\leq A\leq2.2\mu_{0}$, we has a $n(k,t) \propto k^{-3.1(1)}$ over a $k$ range of $10\mu \mathrm{m^{-1}}\leq k \leq 17 \mu \mathrm{m^{-1}}$ and within the time window $20\,\mathrm{ms} \leq t \leq 70\,\mathrm{ms}$, which is shown in Fig. \ref{fig:PowerLaw_nkt}. 

If our system is now a turbulent cloud, then the self-similarity state, a feature of universal behavior, is expected to occur. To obtain the values of universal scaling exponents, we used the likelihood standard procedure in Ref. \cite{Erne2018}. We minimized the function $\chi^{2}(\alpha,\beta)$, which is defined as,

\begin{equation}
    \chi^{2}(\alpha,\beta)=\frac{1}{N^{2}_{t}}\sum_{t=t_{1}}^{t_{N_{t}}}\sum_{t_{0}=t_{1}}^{t_{N_{t}}} \chi^{2}_{\alpha,\beta}(t,t_{0}),
    \label{eq:chi}
\end{equation}

\noindent where we average both times. $t$ is the holding time after the excitation and reference time $t_{0}$ over the $N_{t}$ holding times $\left\{t_{1},\cdots,t_{N_{t}}\right\}$. The functions $\chi^{2}_{\alpha,\beta}(t,t_{0})$ are given by,

\begin{equation}
   \int_{k_{i}}^{k_{f}}dk\frac{\left\{(t/t_{0})^{\alpha}n\left[(t/t_{0})^{\beta}k,t_{0}\right]-n(k,t)\right\}^{2}}{\sigma\left[(t/t_{0})^{\beta}k,t_{0}) \right]^{2}+\sigma(k,t)^{2}},
    \label{eq:chi_int}
\end{equation}

\noindent where $\sigma$ corresponds to the standard deviation of the mean, the quantities $n(k,t)$ and $\sigma(k,t)$ are normalized by the total number of atoms $N(t)$, $n(k,t)=\tilde{n}(k,t)/N(t)$ and $\sigma(k,t)=\tilde\sigma(k,t)/N(t)$.

The limit of integration in Eq. \ref{eq:chi_int} is over the $k$ range $[k_{i}, k_{f}]$. The value of $k_{i}$ is the lowest available momentum from the experimental data, and $k_{f}$ was varied to ensure the results are independent of our choice. For this experiment in particular, we used a $k_{f}=10\,\mu \mathrm{m^{-1}}$ with a variation of $\delta k \approx 0.5\,\mu \mathrm{m^{-1}}$ and $\delta t\approx 5\,\mathrm{ms}$, this is the variation between momentum distribution data.

\begin{figure*}[h!]
    \centering
    \includegraphics[width=0.83\linewidth]{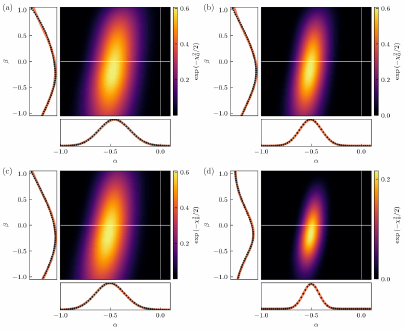}
    \caption{Likelihood functions $L_{A}(\alpha,\beta)$ for three different excitation amplitudes, $1.8\mu_{0}$ (a), $2.0\mu_{0}$ (b), and $2.2\mu_{0}$ (c), and the product of all likelihood functions (d). For panels (a)–(c) we divide the function $\chi^{2}(\alpha,\beta)$ by its minimum value, which we denote by $\chi^{2}(\alpha,\beta)$, so that the maximum value of $L_{A}(\alpha,\beta)$ is the same for all excitation amplitudes. Reproduced from Ref.~\cite{GarciaOrozcoPRA2022}.}
    \label{fig:likelihood}
\end{figure*}

The values of $\alpha$ and $\beta$ are estimated using the likelihood function,

\begin{equation}
    L_{A}(\alpha,\beta)=\exp\left[ -\frac{1}{2}\chi^{2}(\alpha,\beta) \right],
    \label{eq:likelihood_function}
\end{equation}

\noindent where the subscript A corresponds to the label for the different excitation amplitudes that show in Fig. \ref{fig:likelihood}, which have the likelihood function for their different excitation amplitudes, $A=1.8,2.0,\,\mathrm{and}\,2.2\mu_{0}$, respectively.

The value of exponents $\alpha$ and $\beta$, and their uncertainties, are extracted from a Gaussian fit of the marginal-likelihood functions (Fig. \ref{fig:likelihood}),

\begin{align}
    L_{\alpha,A}(\alpha) & = \int d\beta L_{A}(\alpha,\beta), \\
    L_{\beta,A}(\beta) & = \int d\alpha L_{A}(\alpha,\beta),
    \label{eq:l_alpha_beta}
\end{align}

\noindent which are presented in the bottom and left panels around the plots in Fig. \ref{fig:likelihood}\textcolor{blue}{(a)}--\ref{fig:likelihood}\textcolor{blue}{(c)}. Since these amplitudes produced essentially the same exponents, the values of $\alpha$ and $\beta$ and their uncertainties are estimated from the combined likelihood function,

\begin{equation}
    L(\alpha,\beta)=\prod_{A}L_{A}(\alpha,\beta),
    \label{eq:l_prod}
\end{equation}

\noindent shown in Fig. \ref{fig:likelihood}\textcolor{blue}{(d)}. 

\begin{figure*}[h]
    \centering
    \includegraphics[width=0.8\linewidth]{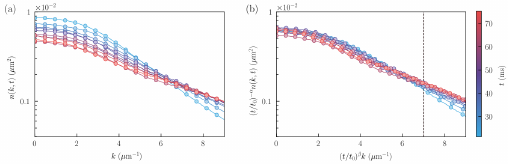}
    \caption{(a) Momentum distributions of a turbulent state for an excitation amplitude $A = 2.2\mu_0$ at different holding times $t$. (b) Corresponding rescaled distributions collapse into a single universal curve, characteristic of self-similar dynamics in the turbulent regime. The universal exponents that described beaviour are $\alpha = -0.50(8)$ and $\beta = -0.2(4)$. The vertical dashed line indicates the infrared cutoff $k_{s}$, delimiting the momentum range where the scaling described by Eq. \ref{eq:scaling} holds. Reproduced from Ref.~\cite{GarciaOrozcoPRA2022}.}
    \label{fig:nk_scaling}
\end{figure*}

The momentum distribution obtained displays universal scaling, meaning that all momentum distributions $n(k,t)$ collapse into a single distribution $n(k,t_{0})$. This behavior indicates that the dynamics can be described by a single scaling function (Eq. \ref{eq:scaling}), and is characteristic of a cloud near an NTFP, as shown in Fig. \ref{fig:nk_scaling}.

\begin{figure}[h]
    \centering
    \includegraphics[width=0.85\linewidth]{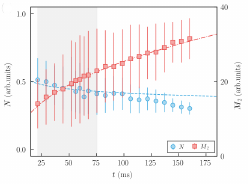}
    \caption{Global quantities, as defined in Eqs. \ref{eq:n} and \ref{eq:m2}, for the excitation amplitudes of $A = 1.8\mu_{0}$. The shaded region corresponds to the time window in which universal behavior is observed. The left axis shows the total number of particles, $\overline{N}$, and the universality range $(k_{s} \leq 7 \mu\mathrm{m^{-1}})$ as a function of $t$. The mean kinetic energy $\overline{M}_{2}$ (right axis) increases with time within the same universal range. The curves follow the theoretical predictions, namely, $\overline{N} \propto t^{\alpha-2\beta}$ and $\overline{M}_{2}\propto t^{-2\beta}$. Reproduced from Ref.~\cite{GarciaOrozcoPRA2022}.}
    \label{fig:N_M2}
\end{figure}

This self-similar evolution in the infrared momentum range is related to particle transport in our closed system over the selected time window. In this condition, two global quantities can be defined, $\overline{N}$ (Eq. \ref{eq:n}) and $\overline{M}_{2}$ (Eq. \ref{eq:m2}). That in our situation, the number of particles over that scaling range $(k_{i}\leq k\leq k_{f})$ is conserved, but the average kinetic energy increases over the same range, as shown in Fig \ref{fig:N_M2}. The results show that the turbulent state exhibits features such as a power-law momentum distribution and self-similar transport between momentum scales, which are characteristic of both quantum turbulence and NTFP dynamics. The values of the exponents $\alpha=-0.50(8)$ and $\beta=-0.2(4)$ show that our system has a universal dynamic behavior and support the classification of the turbulent Bose gases with a system out of equilibrium \cite{Madeira2022Review}, in which energy is transferred from low to high momentum, and this can be confirmed with the sign of alpha. Also, the two global quantities, number of particles $\overline{N}$ (Eq. \ref{eq:n}) and internal energy $\overline{M}_{2}$ (Eq. \ref{eq:m2}) coincide with the exponent values and dimension os the system within the scaling region, for $\overline{N}\propto t^{\alpha-2\beta}$ and $\overline{M}_{2} \propto t^{-2\beta}$. 

\section{Differential equation approach}\label{sec:diferencial_equation}
In Ref. \cite{Madeira2024PNAS}, we introduced a differential equation framework in which the universal scaling associated with NTFPs (Eq. \ref{eq:scaling}) is a solution. This formulation provides a physical interpretation of the scaling exponents associated with the temporal evolution of the amplitude and the rescaling momentum.

The standard method for identifying the presence of NTFPs is usually to verify that numerical or experimental data exhibit the universal scaling of Eq. \ref{eq:scaling}. The standard method is necessary for the momentum distribution in a specific time interval and momentum range and the value of two exponents $\alpha$ and $\beta$ to make all momentum distribution collapse into a single universal function $F(k)$.

However, it is possible to obtain a good estimation of the values of $\alpha$ and $\beta$ using another approach. In this situation, if we consider the momentum distribution as a function of time, $k=k(t)$, the temporal derivative for $n(k,t)$ may be expressed as,

\begin{equation}
    \frac{dn}{dt}=\frac{\partial n}{\partial t}+\frac{\partial n}{\partial k}\frac{dk}{dt}.
    \label{eq:partial_dn_dt}
\end{equation}

To make the last equation allow both amplitude and momentum scale variations to be observed in a self-similar scaling. We consider a particular form for these time dependencies, power laws in time,

\begin{equation}
    \frac{dn}{n}=\alpha \frac{dt}{t},
    \label{eq:df_alpha}
\end{equation}

\begin{equation}
    \frac{dk}{k}=-\beta\frac{dt}{t},
    \label{eq:dif_beta}
\end{equation}

\noindent the exponent $\alpha$ scales the amplitude of the momentum distribution, and the exponent $\beta$ scales $k$. In the limit that the system's evolution is slow, the values of $\alpha$ and $\beta$ can be approximated by zero $(\alpha=\beta=0)$, corresponding to a stationary state, indicating that the system is in equilibrium.

With these assumptions, we can rewrite $n(k,t)$ (Eq. \ref{eq:partial_dn_dt}) obeying the partial differential equation:

\begin{equation}
    t\frac{\partial n(k,t)}{\partial t}=\alpha n(k,t) + \beta k\frac{\partial n(k,t)}{\partial k}.
    \label{eq:n_partial_dif}
\end{equation}

The last equation captures the interplay between temporal changes in amplitude and momentum scale, accounting for the power-law behavior (Eqs. \ref{eq:df_alpha} and \ref{eq:dif_beta}) exhibited by both quantities. If the values of $\alpha$ and $\beta$ are constants, the solution is Eq. \ref{eq:scaling}, which can be verified by direct substitution. A central advantage of our approach is that the universal exponents, $\alpha$ and $\beta$, can be obtained from a limited region of the total momentum distribution, without requiring the entire distribution to collapse into a single function. 

We can consider two limiting scenarios. The first case concerns momentum ranges in which the amplitude of the momentum distribution varies negligibly over time, essentially characterizing regions of minimal change. So, the term $\partial n/\partial t$ can be neglected  ($\partial n/\partial t=0$) and Eq. \ref{eq:n_partial_dif} can be rewritten as,

\begin{equation}
    \left. \frac{\partial n}{\partial k} \right|_{t}=-\frac{\alpha}{\beta}\frac{n}{k},
    \label{eq:nk_partial_tlimit}
\end{equation}

\noindent for this specific situation the solution is,

\begin{equation}
    n(k,t)\propto k^{-\alpha/\beta}
    \label{eq:solution_nk_partial_tlimit}
\end{equation}

\noindent if $n$ is time independet, thus exhibiting a power-law behavior within this specific region. We define $k^{*}$ as the value of k at which the momentum distribution across different holding times is constant. The value of the momentum $k^{*}$ can be found in another theory, such as wave turbulence \cite{Nazarenko2011Wave}. Svistunov \cite{Svistunov1991} found turbulent cascades as a result of $\partial n/\partial t=0$ for a fixed momentum, and this result is similar to Eq. \ref{eq:solution_nk_partial_tlimit}.

\begin{figure}[h!]
    \centering
    \includegraphics[width=0.8\linewidth]{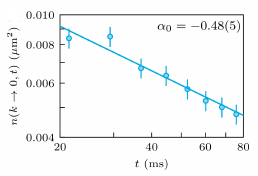}
    \caption{Low-momentum behavior of the momentum distribution as a function of time for a turbulent BEC. The solid curve corresponds to a fit to the functional form provided by Eq. \ref{eq:nk_limit_zero}, which yields $\alpha$ value in agreement with the ones reported by the author Ref.~\cite{GarciaOrozcoPRA2022}. Reproduced from Ref.~\cite{Madeira2024PNAS}.}
    \label{fig:nk_lowmomenta}
\end{figure}

The second interesting case is to take a limit of $ k \xrightarrow{} 0$, deep in the infrared region. In this limit, Eq. \ref{eq:n_partial_dif} can be found that,

\begin{equation}
    n(k\rightarrow 0 ,t) \propto t^{\alpha},
    \label{eq:nk_limit_zero}
\end{equation}

\noindent consistent with taking the same limit directly in Eq. \ref{eq:scaling}. The principal advantage of Eq. \ref{eq:nk_limit_zero} is that it makes tracking the temporal evolution of these nonthermal states easier and more practical. Tracking the evolution of momentum distribution, focusing on the low momenta, allows us to obtain $\alpha$. In Fig. \ref{fig:nk_lowmomenta} shows $n(k\rightarrow0,t)$ as a function of holding time $t$ for the data presented in Ref. \cite{GarciaOrozcoPRA2022}. 

Having determined $\alpha$ from the $k\rightarrow0$ behavior of the distribution, now we can determine $\beta$ using Eq. \ref{eq:n_partial_dif}, evaluation this function around $k^{*}$, 

\begin{equation}
    \beta=-\frac{\alpha}{k^{*}}\frac{n(k^{*},t)}{(\partial n/\partial k)|_{k=k^{*}}}.
    \label{eq:beta_kzero}
\end{equation}

Specifically, we showed that the $\alpha$ can be obtained from the low-momentum limit, $k\xrightarrow{}0$. At the same time, $\beta$ can be determined from regions where the distribution is approximately time-independent, $k\xrightarrow{}k^{*}$.

Our formulation predicts power-law behavior (Eqs. \ref{eq:solution_nk_partial_tlimit} and \ref{eq:nk_limit_zero}), with the exponent given by the ratio of the universal exponents. We validate our approach using data from other experiments, demonstrating its robustness and universal applicability, as shown in Tables \ref{tab:exponents_alpha} and \ref{tab:exponents_beta} for $\alpha$ and $\beta$, respectively.

\begin{table}[h]
\centering
\begin{tabular}{cccc}
\toprule
 && Std. proc. & $k \to 0$ lim. \\
\multicolumn{2}{c}{Experiment} & $\alpha$ & $\alpha_0$ \\
\midrule
&A ($\mu_0$) & & \\
&1.8 & $-0.46(2)$ & $-0.44(5)$  \\
Ref.\cite{GarciaOrozcoPRA2022} & 2.0 & $-0.51(1)$ & $-0.48(5)$ \\
&2.2 & $-0.50(2)$ & $-0.49(5)$ \\
\multicolumn{2}{c}{Ref.\cite{Glidden2021}} & \hspace{0.9em}$1.15(8)$ & \hspace{0.9em}$1.06(5)$ \\
\multicolumn{2}{c}{Ref.\cite{Prufer2018}} & \hspace{0.9em}$0.33(8)$ & \hspace{0.9em}$0.41(8)$ \\
\botrule
\end{tabular}
\caption{The columns \textit{``standard procedure''} correspond to the $\alpha$ values for reports by the authors of Refs. \cite{GarciaOrozcoPRA2022} \cite{Glidden2021} and \cite{Prufer2018}. The $\alpha$ values obtained \textit{``$k\to 0$ limit''} correspond to the low-momentum approximation.}
\label{tab:exponents_alpha}
\end{table}

\begin{table}[h]
\centering
\begin{tabular}{cccc}
\toprule
 && Std. proc. & $k\to k^*$ with $\alpha_0$ \\
\multicolumn{2}{c}{System} & $\beta$ & $\beta_0$ \\
\midrule
&A ($\mu_0$) & &\\
&1.8 & $-0.2(9)$ & $-0.26(7)$ \\
Ref.\cite{GarciaOrozcoPRA2022} & 2.0& $-0.2(7)$ & $-0.25(5)$ \\
&2.2 & $-0.2(9)$ & $-0.37(8)$ \\
\multicolumn{2}{c}{Ref.\cite{Glidden2021}} & \hspace{0.9em}$0.34(5)$ & \hspace{0.9em}$0.41(7)$ \\
\multicolumn{2}{c}{Ref.\cite{Prufer2018}} & \hspace{0.9em}$0.54(6)$ & \hspace{0.9em}$0.57(9)$ \\
\botrule
\end{tabular}
\caption{The columns \textit{``standard procedure''} correspond to the $\beta$ values for reports by the authors of Refs. \cite{GarciaOrozcoPRA2022} \cite{Glidden2021} and \cite{Prufer2018}. The $\beta$ values are obtained using the $\alpha$ value from the last table; the method is described in the Ref. \cite{Madeira2024PNAS} and Eq. \ref{eq:beta_kzero}.}
\label{tab:exponents_beta}
\end{table}

The results show that the differential equation approach provides a powerful and practical framework for analyzing universal dynamics near NTFPs. By focusing on a specific region of the momentum distribution, such as near zero momentum and at the point of minimal temporal variation, it is possible to calculate the universal exponent without relying on full data collapse. The agreement with previously reported experimental and theoretical results \cite{GarciaOrozcoPRA2022,Glidden2021,Prufer2018} confirms the consistency and robustness of this method in different physical systems.

Additionally, the framework links universal scaling to transport processes, predicting power-law behavior in regions of weak temporal evolution and establishes a direct connection between NTFPs and turbulent cascades in far-from-equilibrium dynamics.

\section{Implications for turbulence and transport}\label{sec:transport_turbulence}
\begin{figure*}[h]
    \centering
    \includegraphics[width=0.95\linewidth]{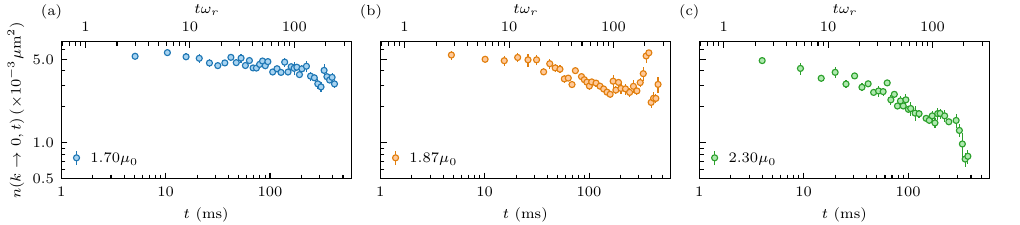}
    \caption{Time evolution of $n(k\xrightarrow{}0,t)$ for three excitation amplitudes. (a) for $A=1.7\mu_{0}$, the system shows a gradual decrease in $n(k\xrightarrow{}0,t)$, indicating a transfer of particles from low to high momentum. (b) for $A=1.87\mu_{0}$, shows a region where $n(k\rightarrow0,t)$ is constant and after some time, a repopulation of the low momentum around $t\approx 220\,\mathrm{ms}$, indicating an inverse cascade of particles. (c) for a large excitation amplitude, such as $A=2.30\mu_{0}$, the repopulation is strongly suppressed. Error bars correspond to one standard deviation. Reproduced from Ref.~\cite{MorenoArmijosPRL2025}.}
    \label{fig:momentum_k0}
\end{figure*}

During the development of a turbulent state, our observations \cite{MorenoArmijosPRL2025} evidence that the system passes through distinct stages in its dynamics. We explored the relaxation dynamics of the turbulent BEC experimentally and identified distinct stages associated with particle and energy transport. We observed that the system transfers particles from low to high momentum, indicating a direct cascade. This direct cascade leads to condensate depletion and exhibits universal dynamical scaling in the system. In Fig. \ref{fig:momentum_k0}, we can observe the evolution of the momentum distribution near $k\rightarrow0$ for three different excitation amplitudes.

In the first situation in Fig. \ref{fig:momentum_k0}\textcolor{blue}{(a)}, the low-momentum states are gradually decreasing, indicating a transfer of particles from low to high momenta. When the excitation is slightly increased, Fig. \ref{fig:momentum_k0}\textcolor{blue}{(b)}, the system promotes the transfer of more atoms from low to high momenta. However, we also observed the inverse process: a repopulation of low-momentum states, indicating a repopulation of the condensate. Nevertheless, if the amplitude is higher as in Fig. \ref{fig:momentum_k0}\textcolor{blue}{(c)}, the condensate depletes faster due to more energy injected into the system; in this case, it does not repopulate.

\begin{figure}[h!]
    \centering
    \includegraphics[width=0.95\linewidth]{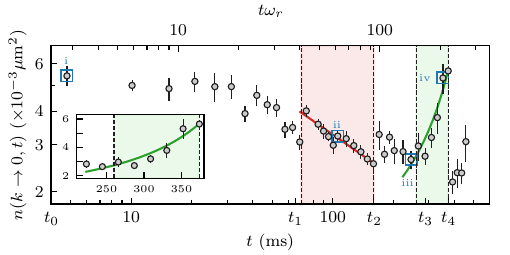}
    \caption{Low-momentum evolution of $n(k \to 0, t)$ for $A = 1.87\mu_0$. After excitation ends $(t_{0}=0)$, the system shows: (i) a slow decrease up to $t_{1}=70\,\text{ms}$ due to direct particle flow to higher momenta; (ii) universal scaling near an NTFP for $t_{1}\to t_{2} =160\,\text{ms}$, $n \propto t^{\alpha}$; (iii) a transient/prethermal stage up to $t_{3}=260\,\text{ms}$; and (iv) an inverse cascade peaking at $t_{4}=370\,\text{ms}$, indicating condensate repopulation. The green curve shows the WTT prediction $n \propto (t_{b}-t)^{\lambda}$ with $\lambda = -1.46$. Error bars denote one standard deviation. Reproduced from Ref.~\cite{MorenoArmijosPRL2025}.}
    \label{fig:nk0_holdtime}
\end{figure}

In terms of amplitude, for $A=1.87\mu_{0}$ (Fig. \ref{fig:nk0_holdtime}), after finishing the excitation at $t_{0}=0$ to $t_{1}=70\,\mathrm{ms}$, the system shows a slow depletion of the particles in the low momenta $n(k\rightarrow0,t)$. In the time interval from $t_{1}$ to $t_{2}=160\,\mathrm{ms}$, the system is near an NTFP and thus exhibits universal scaling. In this range, the system transfers particles from low to high, indicating a direct cascade, with $\alpha =-0.5(1)$ and $\beta=-0.25(7)$. After the direct cascade, over the time interval from $t_{2}$ to $t_{3}=260\,\mathrm{ms}$, we observe a region of little variation that can be associated with a pre-thermalization stage. Then an inverse cascade of particles is observed from $t_{3}$ to $t_{4}=370\,\mathrm{ms}$, so that the peak is in $t_{4}$. 

The $\alpha$ and $\beta$ are the universal exponents associated with a direct energy cascade (shadow red region) during transfer from the low-- to high--momentum region, which depletes the condensate. By controlling the amount of energy given to the cloud, it is possible to observe that after the direct cascade, there is the appearance of an inverse cascade (shadow green region), which indicates a repopulation of low-momentum particles, i.e., the condensate, which can be described using the following equation for the dynamical scaling,

\begin{equation}
    n(k,t)=\tau^{\lambda}n(\tau^{\mu}k,t_{\mathrm{ref}}),
    \label{eq:scaling_tau}
\end{equation}

\noindent where $\tau=(t_{b}-t)/(t_{b}-t_{\mathrm{ref}})$, $t_{b}$ is where the condensate population diverges, $t_{\mathrm{ref}}$ is the same $t_{0}$ and the universal exponents are $\mu$ and $\lambda<0$. Making the differential approach for Eq. \ref{eq:scaling_tau} can yield a similar result in terms of the exponents $\mu$ and $\lambda$, like as,

\begin{equation}
    n(k\rightarrow0,t)\propto(t_{b}-t)^{\lambda},
    \label{eq:scaling_lambda_kzero}
\end{equation}

\begin{figure*}[t]
    \centering
    \includegraphics[width=0.8\linewidth]{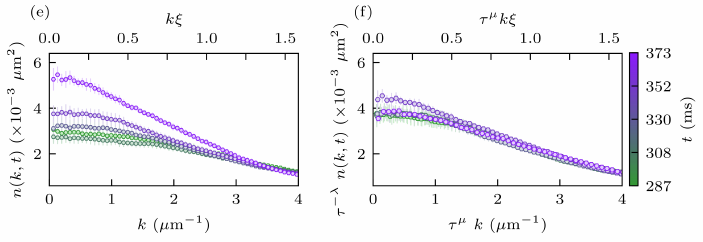}
    \caption{Momentum distribution for the inverse cascade. The momentum distributions (e) for different holding times collapse to a single function after rescaling (f). The behavior shows a repopulation of the low-momentum states, indicating an increase in the particle number in the condensate. The values used for the scaling are $\lambda=-1.5(5)$ and $\mu=-0.9(3)$, which are obtained from the power-law in Eq. \ref{eq:scaling_lambda_kzero}. The error bars correspond to one standard deviation. Reproduced from Ref.~\cite{MorenoArmijosPRL2025}.}
    \label{fig:blowup_nk}
\end{figure*}

For this last region, we obtained values of  $\lambda=-1.5(5)$ and $\mu=-0.9(3)$ using the power-law Eq. \ref{eq:scaling_lambda_kzero}, respectively, for the universal scaling. Fig. \ref{fig:blowup_nk} shows the momentum distribution for the last region (shadow green) and the rescaled momentum distribution using the exponent values $\lambda$ and $\mu$ obtained from Eq. \ref{eq:scaling_lambda_kzero}.

Both processes, direct and inverse cascade, are consistent with the self-similarity solution predicted by wave turbulence theory \cite{Nazarenko2011Wave}, in which nonlinear wave interactions govern the system's behavior out of equilibrium.

These observations show that the relaxation dynamics are governed by the interplay between direct and inverse particle cascades, which describe particle and energy transport across momentum scales. The direct cascade depletes the condensate while exhibiting universal scaling consistent with NTFPs and wave turbulence theory. Under suitable conditions, this process is followed by an inverse cascade that restores the low-momentum population and exhibits distinct self-similar scaling. The agreement between the experimentally obtained exponents and theoretical predictions reinforces the interpretation of turbulence as a universal mechanism driving the system toward equilibrium. Overall, our results demonstrate that multiple scaling regimes can emerge during relaxation, highlighting the central role of turbulence and transport in far-from-equilibrium quantum systems.

\section{Outlook}\label{sec:outlook}
The results discussed in this review show that turbulent Bose gases provide a controllable platform to investigate universal dynamics in far--from--equilibrium quantum systems. In these conditions, the observation of self--similar scaling, together with the identification of nonthermal fixed points and the presence of direct and inverse cascades, indicates that turbulence has an important role in the relaxation dynamics. These results show a direct connection between transport processes in momentum space and universal behavior, reinforcing the interpretation that the system's evolution toward equilibrium.

Despite these advances, several open questions remain. On the theoretical side, a complete description that links WWT in the ultraviolet regime to the strongly occupied infrared regime is still lacking. In this situation, the role of vortices, coherence, and nonlinear interactions in determining the scaling behavior across different momentum ranges needs further investigation. On the experimental side, improvements in momentum resolution and access to larger scaling regions in both time and momentum are important to obtain more precise values of the universal exponents. Additionally, exploring the excitation protocol and initial conditions in our system may enable exploration of different dynamical regimes and identification of universality classes.

The differential equation approach presented in this work provides an alternative means of analyzing universal scaling. In this situation, the method allows the extraction of the universal exponents from limited regions of the momentum distribution, without requiring a full scaling collapse. This approach can be extended to other systems and experimental configurations, providing a useful tool to study nonequilibrium dynamics. In general, these results indicate that the study of turbulent quantum gases will continue to contribute to the understanding of transport processes and approaches to thermalization in quantum many-body systems.

\section*{Funding}
The authors thank M.A.~Caracanhas and L.~Madeira for fruitful discussions. This work was supported by the São Paulo Research Foundation (FAPESP) under grants 2013/07276-1, 2014/50857-8, by the cooperation between FAPESP and the French National Research Agency (ANR) through project RELAQS number 2024/04637-8. The authors also acknowledge the support of the National Council for Scientific and Technological Development (CNPq) under the grants 465360/2014-9. S.S, M.A.M-A, A.D.G-O, and A.F. acknowledge the support from FAPESP -- Finance codes No. 2024/14764-7, No. 2025/13137-1, No. 2025/07547-2, No. 2024/08433-8 and No. 2024/21658-9. V.S.B. acknowledges Texas A\&M University at College Station - TX.

\bibliography{bibliography}

@article{GarciaOrozcoPRA2022,
  title = {Universal dynamics of a turbulent superfluid {B}ose gas},
  author = {Garc\'{\i}a-Orozco, A. D. and Madeira, L. and Moreno-Armijos, M. A. and Fritsch, A. R. and Tavares, P. E. S. and Castilho, P. C. M. and Cidrim, A. and Roati, G. and Bagnato, V. S.},
  journal = {Phys. Rev. A},
  volume = {106},
  issue = {2},
  pages = {023314},
  numpages = {10},
  year = {2022},
  month = {Aug},
  publisher = {American Physical Society},
  doi = {10.1103/PhysRevA.106.023314},
  url = {https://link.aps.org/doi/10.1103/PhysRevA.106.023314}
}

@article{MorenoArmijosPRL2025,
  title = {Observation of Relaxation Stages in a Nonequilibrium Closed Quantum System: Decaying Turbulence in a Trapped Superfluid},
  author = {Moreno-Armijos, M. A. and Fritsch, A. R. and Garc\'{\i}a-Orozco, A. D. and Sab, S. and Telles, G. and Zhu, Y. and Madeira, L. and Nazarenko, S. and Yukalov, V. I. and Bagnato, V. S.},
  journal = {Phys. Rev. Lett.},
  volume = {134},
  issue = {2},
  pages = {023401},
  numpages = {6},
  year = {2025},
  month = {Jan},
  publisher = {American Physical Society},
  doi = {10.1103/PhysRevLett.134.023401},
  url = {https://link.aps.org/doi/10.1103/PhysRevLett.134.023401}
}

@article{ScheppachPRA2010,
  title = {Matter-wave turbulence: Beyond kinetic scaling},
  author = {Scheppach, Christian and Berges, J\"urgen and Gasenzer, Thomas},
  journal = {Phys. Rev. A},
  volume = {81},
  issue = {3},
  pages = {033611},
  numpages = {16},
  year = {2010},
  month = {Mar},
  publisher = {American Physical Society},
  doi = {10.1103/PhysRevA.81.033611},
  url = {https://link.aps.org/doi/10.1103/PhysRevA.81.033611}
}

@article{NowakPRA2012,
  title = {Nonthermal fixed points, vortex statistics, and superfluid turbulence in an ultracold {B}ose gas},
  author = {Nowak, Boris and Schole, Jan and Sexty, D\'enes and Gasenzer, Thomas},
  journal = {Phys. Rev. A},
  volume = {85},
  issue = {4},
  pages = {043627},
  numpages = {19},
  year = {2012},
  month = {Apr},
  publisher = {American Physical Society},
  doi = {10.1103/PhysRevA.85.043627},
  url = {https://link.aps.org/doi/10.1103/PhysRevA.85.043627}
}

@article{BergesPRL2008,
  title = {Nonthermal Fixed Points: Effective Weak Coupling for Strongly Correlated Systems Far from Equilibrium},
  author = {Berges, J\"urgen and Rothkopf, Alexander and Schmidt, Jonas},
  journal = {Phys. Rev. Lett.},
  volume = {101},
  issue = {4},
  pages = {041603},
  numpages = {4},
  year = {2008},
  month = {Jul},
  publisher = {American Physical Society},
  doi = {10.1103/PhysRevLett.101.041603},
  url = {https://link.aps.org/doi/10.1103/PhysRevLett.101.041603}
}

@article{BergesNuclearPhysicsB2009,
  title = {Nonthermal fixed points and the functional renormalization group},
  journal = {Nuclear Physics B},
  volume = {813},
  number = {3},
  pages = {383-407},
  year = {2009},
  issn = {0550-3213},
  doi = {https://doi.org/10.1016/j.nuclphysb.2008.12.017},
  url = {https://www.sciencedirect.com/science/article/pii/S0550321308007219},
  author = {Jürgen Berges and Gabriele Hoffmeister}
}

@article{BergesPRD2011,
  title = {Strong versus weak wave-turbulence in relativistic field theory},
  author = {Berges, J\"urgen and Sexty, D\'enes},
  journal = {Phys. Rev. D},
  volume = {83},
  issue = {8},
  pages = {085004},
  numpages = {8},
  year = {2011},
  month = {Apr},
  publisher = {American Physical Society},
  doi = {10.1103/PhysRevD.83.085004},
  url = {https://link.aps.org/doi/10.1103/PhysRevD.83.085004}
}

@article{DosSantosFilhoPhysicsD2022,
  title = {Incompressible energy spectrum from wave turbulence},
  journal = {Physica D: Nonlinear Phenomena},
  volume = {440},
  pages = {133479},
  year = {2022},
  issn = {0167-2789},
  doi = {https://doi.org/10.1016/j.physd.2022.133479},
  url = {https://www.sciencedirect.com/science/article/pii/S0167278922001981},
  author = {Marcos A.G. {dos Santos Filho} and Francisco E.A. {dos Santos}}
}

@misc{Zakharov1992,
  title={Kolmogorov Spectra of Turbulence I},
  author={L’vov, VE Zakharov VS and Falkovich, G},
  year={1992},
  publisher={Springer Berlin, Heidelberg}
}

@misc{Nazarenko2011Wave,
  title={Wave turbulence},
  author={Nazarenko, Sergey},
  volume={825},
  year={2011},
  publisher={Springer Science \& Business Media}
}

@incollection{NowakOxford2016,
  author = {Nowak, Boris and Erne, Sebastian and Karl, Markus and Schole, Jan and Sexty, Dénes and Gasenzer, Thomas},
  isbn = {9780198768166},
  title = {Nonthermal fixed points: universality, topology, and turbulence in {B}ose gases},
  booktitle = {Strongly Interacting Quantum Systems out of Equilibrium: Lecture Notes of the Les Houches Summer School: Volume 99, August 2012},
  publisher = {Oxford University Press},
  year = {2016},
  month = {06},
  doi = {10.1093/acprof:oso/9780198768166.003.0007},
  eprint = {https://academic.oup.com/book/0/chapter/192842338/chapter-pdf/44202982/acprof-9780198768166-chapter-7.pdf},
}

@article{Mikheev2023EPJSP,
  title={Universal dynamics and non-thermal fixed points in quantum fluids far from equilibrium},
  author={Mikheev, Aleksandr N and Siovitz, Ido and Gasenzer, Thomas},
  journal={The European Physical Journal Special Topics},
  volume={232},
  number={20},
  pages={3393--3415},
  year={2023},
  publisher={Springer}
}

@article{Erne2018,
  author = {Ern{\'e}, S. and B{\"u}cker, R. and Gasenzer, T. and Berges, J. and Schmiedmayer, J.},
  title = {Universal dynamics in an isolated one-dimensional {B}ose gas far from equilibrium},
  journal = {Nature},
  volume = {563},
  pages = {225--229},
  year = {2018},
  doi = {10.1038/s41586-018-0667-0}
}

@article{Prufer2018,
  author = {Pr{\"u}fer, M. and Kunkel, P. and Strobel, H. and Lannig, S. and Linnemann, D. and Schmied, C. M. and Berges, J. and Gasenzer, T. and Oberthaler, M. K.},
  title = {Observation of universal dynamics in a spinor {B}ose gas far from equilibrium},
  journal = {Nature},
  volume = {563},
  pages = {217--220},
  year = {2018},
  doi = {10.1038/s41586-018-0659-0}
}

@article{Glidden2021,
  author = {Glidden, J. A. P. and Eigen, C. and Dogra, L. H. and Hilker, T. A. and Smith, R. P. and Hadzibabic, Z.},
  title = {Bidirectional dynamic scaling in an isolated {B}ose gas far from equilibrium},
  journal = {Nature Physics},
  volume = {17},
  pages = {457--461},
  year = {2021},
  doi = {10.1038/s41567-020-01114-x}
}

@article{Madeira2022Review,
  author = {Madeira, Lucas and Bagnato, Vanderlei S.},
  title = {Non-Thermal Fixed Points in {B}ose Gas Experiments},
  journal = {Symmetry},
  volume = {14},
  number = {4},
  pages = {678},
  year = {2022},
  doi = {10.3390/sym14040678}
}

@article{Madeira2024PNAS,
  author = {Madeira, Lucas and García-Orozco, Arnol D. and Moreno-Armijos, Michelle A. and Fritsch, Amilson R. and Bagnato, Vanderlei S.},
  title = {Universal scaling in far-from-equilibrium quantum systems: An equivalent differential approach},
  journal = {Proceedings of the National Academy of Sciences},
  volume = {121},
  number = {30},
  pages = {e2404828121},
  year = {2024},
  doi = {10.1073/pnas.2404828121}
}

@article{Thompson2014TurbulenceBEC,
  author = {Thompson, K. J. and Bagnato, G. G. and Telles, G. D. and Caracanhas, M. A. and dos Santos, F. E. A. and Bagnato, V. S.},
  title = {Evidence of power-law behavior in the momentum distribution of a turbulent trapped {B}ose-{E}instein condensate},
  journal = {Laser Physics Letters},
  volume = {11},
  pages = {015501},
  year = {2014},
  doi = {10.1088/1612-2011/11/1/015501}
}

@article{Navon2016TurbulentCascade,
  author = {Navon, N. and Gaunt, A. L. and Smith, R. P. and Hadzibabic, Z.},
  title = {Emergence of a turbulent cascade in a quantum gas},
  journal = {Nature},
  volume = {539},
  pages = {72--75},
  year = {2016},
  doi = {10.1038/nature20114}
}

@article{Henn2009TurbulencePRL,
  author = {Henn, E. A. L. and Seman, J. A. and Roati, G. and Magalh{\~a}es, K. M. F. and Bagnato, V. S.},
  title = {Emergence of turbulence in an oscillating {B}ose-{E}instein condensate},
  journal = {Physical Review Letters},
  volume = {103},
  pages = {045301},
  year = {2009},
  doi = {10.1103/PhysRevLett.103.045301}
}

@article{Henn2009TurbulencePRA,
  author = {Henn, E. A. L. and Seman, J. A. and Ramos, E. R. F. and Caracanhas, M. and Castilho, P. and Ol{\'i}mpio, E. P. and Roati, G. and Magalh{\~a}es, D. V. and Magalh{\~a}es, K. M. F. and Bagnato, V. S.},
  title = {Turbulent regime in a trapped {B}ose-{E}instein condensate},
  journal = {Physical Review A},
  volume = {79},
  pages = {043618},
  year = {2009},
  doi = {10.1103/PhysRevA.79.043618}
}

@article{Madison2001Vortices,
  author = {Madison, K. W. and Chevy, F. and Bretin, V. and Dalibard, J.},
  title = {Vortex formation in a stirred {B}ose-{E}instein condensate},
  journal = {Physical Review Letters},
  volume = {86},
  pages = {4443--4446},
  year = {2001},
  doi = {10.1103/PhysRevLett.86.4443}
}

@article{Berges2015,
  author    = {J. Berges and K. Boguslavski and S. Schlichting and R. Venugopalan},
  title     = {Universal attractor in a highly occupied non-Abelian plasma},
  journal   = {Physical Review Letters},
  volume    = {114},
  pages     = {061601},
  year      = {2015},
  doi       = {10.1103/PhysRevLett.114.061601}
}

@article{Chantesana2019,
  author    = {I. Chantesana and A. P. Orioli and T. Gasenzer},
  title     = {Kinetic theory of nonthermal fixed points in a {B}ose gas},
  journal   = {Physical Review A},
  volume    = {99},
  pages     = {043620},
  year      = {2019},
  doi       = {10.1103/PhysRevA.99.043620}
}

@article{Widom1965,
  author  = {Widom, B.},
  title   = {Equation of state in the neighborhood of the critical point},
  journal = {Journal of Chemical Physics},
  volume  = {43},
  pages   = {3898},
  year    = {1965},
  doi     = {10.1063/1.1696618}
}

@article{Kadanoff1966,
  author  = {Kadanoff, L. P.},
  title   = {Scaling laws for Ising models near Tc},
  journal = {Physics Physique Fizika},
  volume  = {2},
  pages   = {263},
  year    = {1966},
  doi     = {10.1103/PhysicsPhysiqueFizika.2.263}
}

@article{Wilson1971a,
  author  = {Wilson, K. G.},
  title   = {Renormalization group and critical phenomena. I. Renormalization group and the Kadanoff scaling picture},
  journal = {Physical Review B},
  volume  = {4},
  pages   = {3174},
  year    = {1971},
  doi     = {10.1103/PhysRevB.4.3174}
}

@article{Wilson1971b,
  author  = {Wilson, K. G.},
  title   = {Renormalization group and critical phenomena. II. Phase-space cell analysis of critical behavior},
  journal = {Physical Review B},
  volume  = {4},
  pages   = {3184},
  year    = {1971},
  doi     = {10.1103/PhysRevB.4.3184}
}

@article{Hohenberg1977,
  author  = {Hohenberg, P. C. and Halperin, B. I.},
  title   = {Theory of dynamic critical phenomena},
  journal = {Reviews of Modern Physics},
  volume  = {49},
  pages   = {435},
  year    = {1977},
  doi     = {10.1103/RevModPhys.49.435}
}

@incollection{Janssen1979,
  author    = {Janssen, H. K.},
  title     = {Field-theoretic methods applied to critical dynamics},
  booktitle = {Dynamical Critical Phenomena and Related Topics},
  series    = {Lecture Notes in Physics},
  volume    = {104},
  publisher = {Springer},
  address   = {Heidelberg},
  pages     = {26},
  year      = {1979},
  doi       = {10.1007/3-540-09523-3_2}
}

@article{Mikheev2023,
  author  = {Mikheev, A. N. and Siovitz, S. and Gasenzer, T.},
  title   = {Universal dynamics and non-thermal fixed points in quantum fluids far from equilibrium},
  journal = {European Physical Journal Special Topics},
  volume  = {232},
  pages   = {1--28},
  year    = {2023},
  doi     = {10.1140/epjs/s11734-023-00974-7}
}

@article{Schmied2019,
  author  = {Schmied, C.-M. and Mikheev, A. N. and Gasenzer, T.},
  title   = {Bidirectional universal dynamics in a spinor {B}ose gas close to a nonthermal fixed point},
  journal = {Physical Review A},
  volume  = {99},
  pages   = {033611},
  year    = {2019},
  doi     = {10.1103/PhysRevA.99.033611}
}

@phdthesis{MorenoArmijos2025Thesis,
  author    = {Moreno Armijos, Michelle Alejandra},
  title     = {Emergent nonequilibrium dynamics in decaying turbulence of {B}ose-Einstein condensates},
  school    = {Instituto de Física de São Carlos, Universidade de São Paulo},
  address   = {São Carlos, Brazil},
  year      = {2025},
  type      = {PhD Thesis},
  pages     = {118},
  advisor   = {Bagnato, Vanderlei Salvador}
}

@phdthesis{GarciaOrozco2023Thesis,
  author    = {García-Orozco, Arnol Daniel},
  title     = {Turbulent {B}ose-{E}instein condensates as an out-of-equilibrium quantum systems},
  school    = {Instituto de Física de São Carlos, Universidade de São Paulo},
  address   = {São Carlos, Brazil},
  year      = {2023},
  type      = {PhD Thesis},
  pages     = {179},
  advisor   = {Bagnato, Vanderlei Salvador}
}

@article{Svistunov1991,
  author  = {Svistunov, B. V.},
  title   = {Highly nonequilibrium {B}ose condensation in a weakly interacting gas},
  journal = {Journal of the Moscow Physical Society},
  volume  = {1},
  pages   = {373},
  year    = {1991}
}

@article{Neely2013,
  author = {Neely, T. W. and Bradley, A. S. and Samson, E. C. and Rooney, S. J. and Wright, E. M. and Law, K. J. H. and Carretero-Gonz\'alez, R. and Kevrekidis, P. G. and Davis, M. J. and Anderson, B. P.},
  title = {Characteristics of Two-Dimensional Quantum Turbulence in a Compressible Superfluid},
  journal = {Phys. Rev. Lett.},
  volume = {111},
  pages = {235301},
  year = {2013},
  doi = {10.1103/PhysRevLett.111.235301}
}

@article{Weiler2008,
  author = {Weiler, C. N. and Neely, T. W. and Scherer, D. R. and Bradley, A. S. and Davis, M. J. and Anderson, B. P.},
  title = {Spontaneous Vortices in the Formation of Bose-Einstein Condensates},
  journal = {Nature},
  volume = {455},
  pages = {948--951},
  year = {2008},
  doi = {10.1038/nature07334}
}

@article{Khaykovich2002,
  author = {Khaykovich, L. and Schreck, F. and Ferrari, G. and Bourdel, T. and Cubizolles, J. and Carr, L. D. and Castin, Y. and Salomon, C.},
  title = {Formation of a Matter-Wave Bright Soliton},
  journal = {Science},
  volume = {296},
  pages = {1290--1293},
  year = {2002},
  doi = {10.1126/science.1071021}
}

@article{Hung2011,
  author = {Hung, C.-L. and Zhang, X. and Gemelke, N. and Chin, C.},
  title = {Observation of Scale Invariance and Universality in Two-Dimensional Bose Gases},
  journal = {Nature},
  volume = {470},
  pages = {236--239},
  year = {2011},
  doi = {10.1038/nature09722}
}

@article{Makotyn2014,
  author = {Makotyn, P. and Klauss, C. E. and Goldberger, D. L. and Cornell, E. A. and Jin, D. S.},
  title = {Universal Dynamics of a Degenerate Unitary Bose Gas},
  journal = {Nat. Phys.},
  volume = {10},
  pages = {116--119},
  year = {2014},
  doi = {10.1038/nphys2850}
}

@article{Nowak2011,
  author = {Nowak, B. and Sexty, D. and Gasenzer, T.},
  title = {Superfluid Turbulence: Nonthermal Fixed Point in an Ultracold Bose Gas},
  journal = {Phys. Rev. B},
  volume = {84},
  pages = {020506},
  year = {2011},
  doi = {10.1103/PhysRevB.84.020506}
}

@article{Kofman1994,
  author = {Kofman, L. and Linde, A. and Starobinsky, A.},
  title = {Reheating after inflation},
  journal = {Physical Review Letters},
  volume = {73},
  pages = {3195--3198},
  year = {1994}
}

@article{Micha2004,
  author = {Micha, R. and Tkachev, I.},
  title = {Turbulent thermalization},
  journal = {Physical Review D},
  volume = {70},
  pages = {043538},
  year = {2004}
}

@article{Berges2008,
  author = {Berges, J. and Scheffler, S. and Sexty, D.},
  title = {Turbulence in nonabelian gauge theory},
  journal = {Physical Review D},
  volume = {77},
  pages = {034504},
  year = {2008}
}

@article{Allahverdi2010,
  author = {Allahverdi, R. and Brandenberger, R. and Cyr-Racine, F.-Y. and Mazumdar, A.},
  title = {Reheating in inflationary cosmology: Theory and applications},
  journal = {Annual Review of Nuclear and Particle Science},
  volume = {60},
  pages = {27--51},
  year = {2010}
}

@article{Polkovnikov2011,
  author = {Polkovnikov, A. and Sengupta, K. and Silva, A. and Vengalattore, M.},
  title = {Colloquium: Nonequilibrium dynamics of closed interacting quantum systems},
  journal = {Reviews of Modern Physics},
  volume = {83},
  pages = {863--883},
  year = {2011}
}

@article{Calabrese2006,
  author = {Calabrese, P. and Cardy, J.},
  title = {Time dependence of correlation functions following a quantum quench},
  journal = {Physical Review Letters},
  volume = {96},
  pages = {136801},
  year = {2006}
}

@article{Eisert2015,
  author = {Eisert, J. and Friesdorf, M. and Gogolin, C.},
  title = {Quantum many-body systems out of equilibrium},
  journal = {Nature Physics},
  volume = {11},
  pages = {124--130},
  year = {2015}
}

@article{Mitra2018,
  author = {Mitra, A.},
  title = {Quantum quench dynamics},
  journal = {Annual Review of Condensed Matter Physics},
  volume = {9},
  pages = {245--259},
  year = {2018}
}

@article{Deutsch1991,
  author = {Deutsch, J. M.},
  title = {Quantum statistical mechanics in a closed system},
  journal = {Physical Review A},
  volume = {43},
  pages = {2046--2049},
  year = {1991}
}

@article{Srednicki1994,
  author = {Srednicki, M.},
  title = {Chaos and quantum thermalization},
  journal = {Physical Review E},
  volume = {50},
  pages = {888--901},
  year = {1994}
}

@article{Rigol2008,
  author = {Rigol, M. and Dunjko, V. and Olshanii, M.},
  title = {Thermalization and its mechanism for generic isolated quantum systems},
  journal = {Nature},
  volume = {452},
  pages = {854--858},
  year = {2008}
}

@article{Dalessio2016,
  author = {D'Alessio, L. and Kafri, Y. and Polkovnikov, A. and Rigol, M.},
  title = {From quantum chaos and eigenstate thermalization to statistical mechanics},
  journal = {Advances in Physics},
  volume = {65},
  pages = {239--362},
  year = {2016}
}

\end{document}